\documentclass{PoS}

\usepackage{color}
\usepackage{amsmath}

\newcommand{\x}{\ensuremath{\phantom{0}}}
\newcommand{\y}{\ensuremath{\phantom{.}}}
\newcommand{\Sig}{${\rm B}{}^{+}$$\rightarrow$${\rm K{}^{+}}$$\mu^+\mu^-$~}
\newcommand{\JPS}{${\rm B}{}^{+}$$\rightarrow$${\rm K{}^{+}}$$J/\psi(\mu^+\mu^-)$~}
\newcommand{\PsP}{${\rm B}{}^{+}$$\rightarrow$${\rm K{}^{+}}$$\psi'(\mu^+\mu^-)$~}
\newcommand{\bts}{\ensuremath{b \to s \ell^+ \ell^-}~}
\newcommand{\BtoKstmumu}{${\rm B}{}^{0}$$\rightarrow$${\rm K{}^{*0}}$$\mu^+\mu^-$~}

\title{Angular analyses of $b \to s \mu^+ \mu^-$ transitions at CMS}

\ShortTitle{Angular analyses of $b \to s \mu^+ \mu^-$ transitions at CMS}

\author{\speaker{Dayong Wang}\thanks{Partially supported by the
    Ministry of Science and Technology of China, under Grants
    No. 2013CB837800.} \\
  for the CMS Collaboration\\
  \\
  State Key Laboratory of Nuclear Physics and
  Technology, Peking University, Beijing, China\\
  E-mail: \email{dayong.wang@pku.edu.cn}}

      \abstract{

        The flavour changing neutral current decays can be interesting
        probes for searching for new physics.  Angular distributions
        of \bts transition processes of both
        $\mathrm{B}^0 \to \mathrm{K}^{*0} \mu^ +\mu^-$ and \Sig are
        studied using a sample of proton-proton collisions at
        $\sqrt{s} = 8~\mathrm{TeV}$ collected with the CMS detector at
        the LHC, corresponding to an integrated luminosity of
        $20.5~\mathrm{fb}^{-1}$.  Angular analyses are performed to
        determine $P_1$ and $P_5'$ angular parameters for \BtoKstmumu
        and $A_{FB}$ and $F_{H}$ parameters for
        ${\rm B}{}^{+}$$\rightarrow$${\rm
          K{}^{+}}$$\mu^+\mu^-$, all as functions of the dimuon
        invariant mass squared. The
        $P_5'$ parameter is of particular interest due to recent
        measurements that indicate a potential discrepancy with the
        standard model.  All the measurements are consistent with the
        standard model predictions.  Efforts with more channels and
        more coming data will be continued to further test the
        standard model with higher precision in future. }

      \FullConference{The International Conference on B-Physics at Frontier Machines - BEAUTY2018\\
        6-11 May, 2018\\
        La Biodola, Elba Island, Italy}

\begin{document}

\section{Introduction}

The Compact Muon Solenoid (CMS) at CERN~\cite{CMS} is a general-purpose
detector currently running on the large hadron collider (LHC). It is
equipped with large area of silicon trackers, a 3.8T magnetic field,
superb muon detection systems with large acceptance and very flexible
trigger systems. These features make CMS an ideal detector for
performing precise measurements of heavy flavor physics.

\begin{figure}[!htb]
    \centering
	\includegraphics[width=0.29\textwidth]{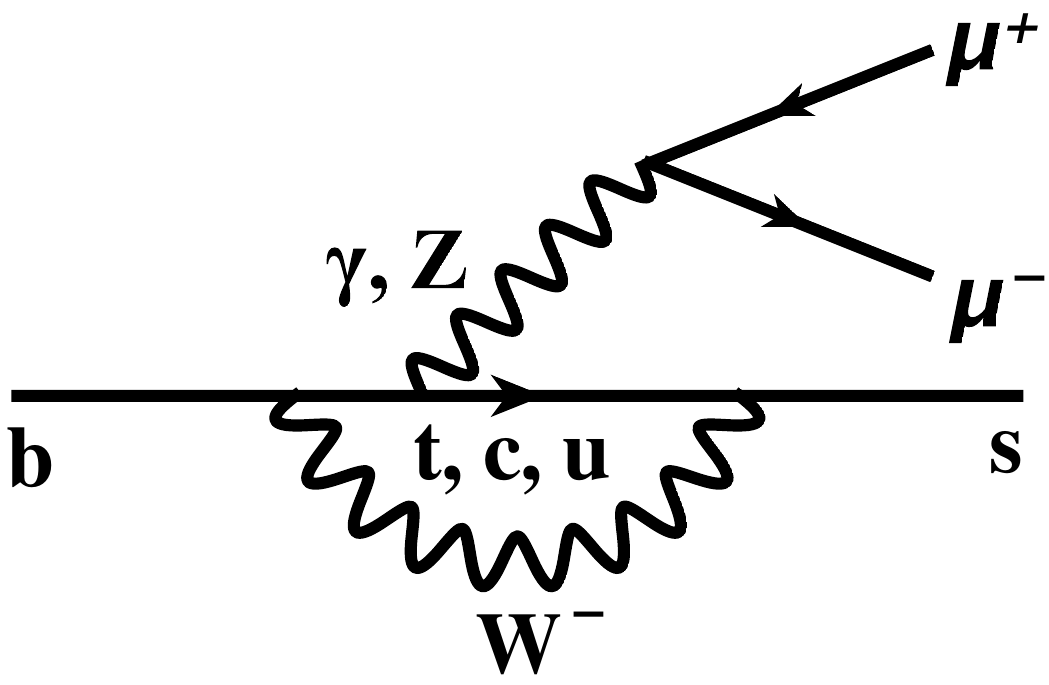}
	\includegraphics[width=0.29\textwidth]{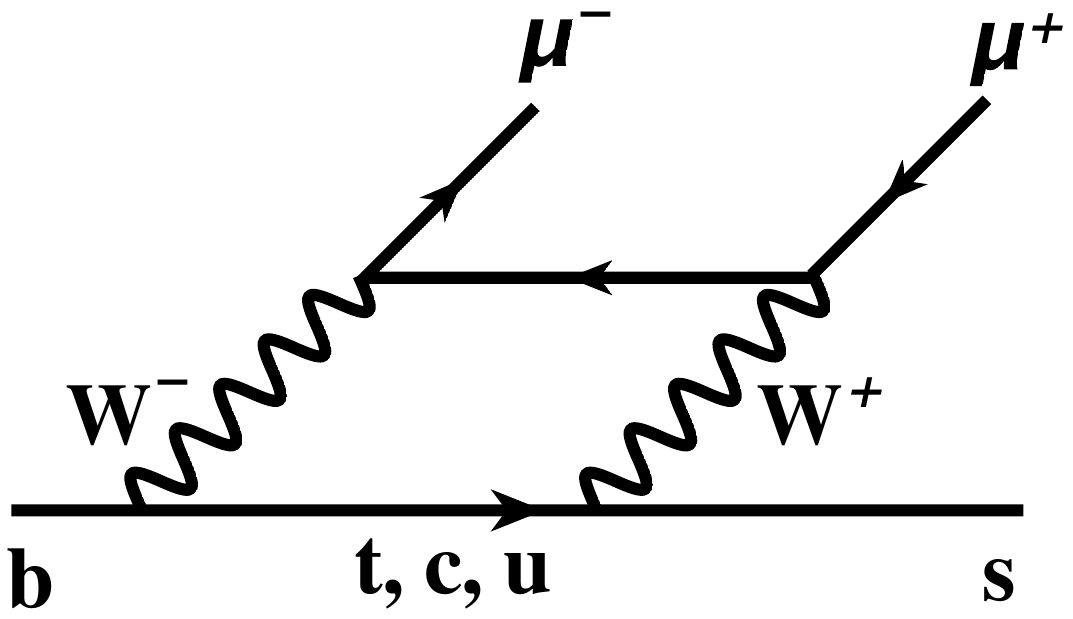}
        \caption{\label{fig-diagrams} The SM electroweak penguin
          (Left) and box (Right) diagrams for the transition \bts. }
\end{figure}

Phenomena beyond the standard model (SM) of particle physics can
become manifest directly, via the production of new particles, or
indirectly, by affecting the production and decay of SM particles.
The transitions of the type \bts is a flavor-changing neutral current
(FCNC) process, with $\ell$ denoting a charged lepton. In the SM, this
type of transition is forbidden at tree level and occurs through
higher-order processes via either electroweak $Z/\gamma$
penguin diagrams or a $W^+W^-$ box diagram, as shown in
Fig.~\ref{fig-diagrams}. This makes the measurement of these rare FCNC
decays more sensitive to possible physics phenomena beyond the SM
(BSM).

CMS has recently analysed two such FCNC decays:
${\rm B}{}^{0}$$\rightarrow$${\rm K{}^{*0}}$$\mu^+\mu^-$, where
$\mathrm{K}^{*0}$ indicates the $\mathrm{K}^{*0}(892)$ meson, and
\Sig. Both analyses use a sample of events collected in proton-proton
(pp) collisions at a center-of-mass energy of $8~\mathrm{TeV}$ with
the CMS detector at LHC. The data correspond to an integrated
luminosity of $20.5~\mathrm{fb}^{-1}$~\cite{LUMI}. The data for these
analysis was recorded using a low-mass dimuon HLT with a displaced
vertex.

\section{Angular analysis of $\mathrm{B}^0 \to \mathrm{K}^{*0} \mu^+\mu^-$~\cite{Sirunyan:2017dhj}
}
\label{sec:formular}

\begin{figure}[!htb]
\centering
\includegraphics[width=0.7\textwidth]{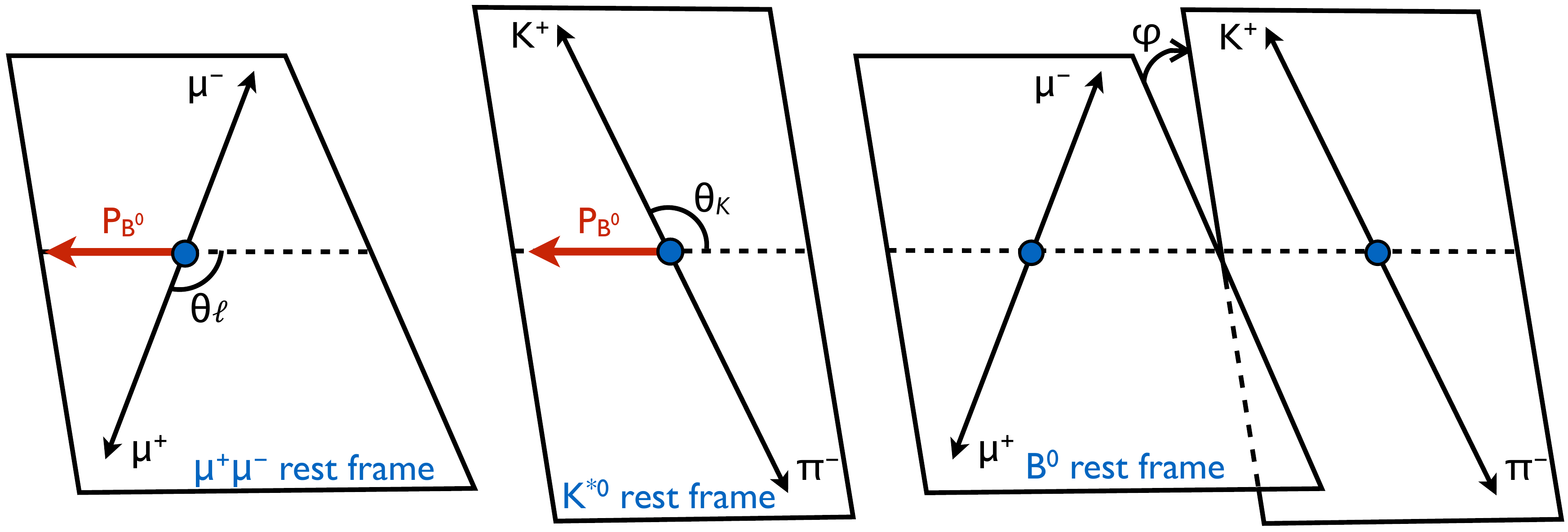}
\caption{ Sketch showing the definition of the angular variables $\theta_\ell$ (left), $\theta_K$ (middle), and $\varphi$ (right) for the decay $\mathrm{B}^0 \to \mathrm{K}^{*0} \mu^ +\mu^-$.}
\label{fig:ske}
\end{figure}

The differential decay rate for
$\mathrm{B}^0 \to \mathrm{K}^{*0} \mu^ +\mu^-$ can be written in terms
of the dimuon mass squared ($q^2$) and three angular variables as a
combination of spherical harmonics. Figure~\ref{fig:ske} shows the
angular variables defining the decay kinematics: $\theta_\ell$ is the
angle between the positive (negative) muon momentum and the direction
opposite to the $\mathrm{B}^0$ ($\bar{\mathrm{B}^0}$) in the dimuon
rest frame, $\theta_K$ is the angle between the kaon momentum and the
direction opposite to the $\mathrm{B}^0$ ($\bar{\mathrm{B}^0}$) in the
$\mathrm{K}^{*0}$ ($\bar{\mathrm{K}^{*0}}$) rest frame, and $\varphi$
is the angle between the plane containing the two muons and the plane
containing the kaon and pion in the $\mathrm{B}^0$ rest frame.

New physics may modify any of the angular variables~\cite{
  DescotesGenon:2012zf} relative to their SM values\cite{Jager:2012uw,
  Descotes-Genon:2013vna}.  Previous measurements of some of
these quantities by the BaBar, Belle, CDF, LHCb, and CMS experiments
are consistent with the SM\cite{CMSKstarmumu2016}. The $P_5'$ parameter
in the decay $\mathrm{B}^0 \to \mathrm{K}^{*0} \mu^ +\mu^-$~ is of
particular interest due to recent LHCb and Belle
measurements~\cite{LHCbP5p1, LHCbP5p2, BelleP5p} that indicate a
potential discrepancy with the standard model. CMS performed a new
measurement of the $P_1$ and $P_5'$ angular parameters
\cite{Sirunyan:2017dhj}, trying to elucidate the situation. In the
measurement, the values of $P_1$ and $P_5'$ angular parameters are
determined by fitting the distribution of events as a function of the
three angular variables. All measurements are
performed in $q^2$ bins from 1 to $19~\mathrm{GeV}^2$. The $q^2$ bins
$8.68<q^2<10.09~\mathrm{GeV}^2$ and $12.90<q^2<14.18~\mathrm{GeV}^2$, corresponding
to $\mathrm{B}^0 \to \mathrm{K}^{*0} J/\psi$ and
$\mathrm{B}^0 \to \mathrm{K}^{*0} \mathrm{\psi}^{'}$ decays,
respectively, are used to validate the analysis.

There can be a contribution from spinless (S-wave)
$\mathrm{K}^{-}\mathrm{\pi}^{+}$
combinations~\cite{Descotes-Genon:2013vna}. This is parametrized with
three terms: $F_\mathrm{S}$, which is related to the S-wave fraction,
and $A_\mathrm{S}$ and $A^5_\mathrm{S}$, which are the interference
amplitudes between the S-wave and P-wave decays. Including these
components, the angular distribution of
$\mathrm{B}^0 \to \mathrm{K}^{*0} \mu^ +\mu^-$ can be written
as~\cite{Descotes-Genon:2013vna}:

\begin{equation} \label{eq:PDF}
 \begin{split}
   \frac{1} {\mathrm{d} \Gamma / \mathrm{d}q^2} \frac{\mathrm{d}^4\Gamma} {\mathrm{d}q^2 \mathrm{d}\cos\theta_\ell \mathrm{d}\cos\theta_\mathrm{K} \mathrm{d}\varphi} =
 & \frac{9} {8\pi} \left\{ \frac{2}{3} \left[ (F_\mathrm{S}+A_\mathrm{S}\cos\theta_\mathrm{K}) \left( 1-\cos^2\theta_\ell \right) + A^5_\mathrm{S} \sqrt{1-\cos^2\theta_\mathrm{K}} \right. \right. \\
 & \left. \sqrt{1-\cos^2\theta_\ell}\cos\varphi \right] + \left(1 - F_\mathrm{S} \right) \Bigl[ 2F_\mathrm{L}\cos^2\theta_\mathrm{K} \left( 1-\cos^2\theta_\ell \right) \Bigr. \\
 & + \frac{1} {2} \left( 1-F_\mathrm{L} \right) \left( 1-\cos^2\theta_\mathrm{K} \right) \left( 1+\cos^2\theta_\ell \right) + \frac{1} {2} P_1(1-F_\mathrm{L}) \\
 & (1-\cos^2\theta_\mathrm{K})(1-\cos^2\theta_\ell)\cos2 \varphi + 2P_5'\cos\theta_\mathrm{K} \sqrt{F_\mathrm{L} \left( 1-F_\mathrm{L} \right) } \\
 & \Bigl. \left. \sqrt{1-\cos^2\theta_\mathrm{K}} \sqrt{1-\cos^2\theta_\ell}\cos\varphi \Bigr] \right\}.
  \end{split}
\end{equation}
where $F_\mathrm{L}$ denotes the longitudinal polarization fraction of
the $\mathrm{K}^{*0}$. This expression is an exact simplification of
the full angular distribution, obtained by folding the $\varphi$ and
$\theta_\ell$ angles about zero and $\pi/2$,
respectively.

For each $q^2$ bin, the observables of interest are extracted from an
unbinned extended maximum-likelihood fit to four variables: the
$\mathrm{K}^{+}\mathrm{\pi}^{-} \mu^ +\mu^-$ invariant mass $m$ and
the three angular variables ${\theta_\ell}$, ${\theta_K}$, and
$\varphi$. For each $q^2$ bin, the unnormalized probability density
function (pdf) has the following expression:

\begin{equation} \label{eq:angALL}
  \begin{split}
    \mathrm{pdf}(m,\theta_K,\theta_\ell,\varphi) & = Y^{C}_{S} \biggl[ S^{C}(m)  \, S^a(\theta_K,\theta_\ell,\varphi) \, \epsilon^{C}(\theta_K,\theta_\ell,\varphi) \biggr. \\
    & \biggl. + \frac{f^{M}}{1-f^{M}}~S^{M}(m) \, S^a(-\theta_K,-\theta_\ell,\varphi) \, \epsilon^{M}(\theta_K,\theta_\ell,\varphi) \biggr] \\
    & + Y_{B}\,B^m(m) \, B^{\theta_K}(\theta_K) \, B^{\theta_\ell}(\theta_\ell) \, B^{\varphi}(\varphi), \\
  \end{split}
\end{equation}
where the contributions correspond to correctly tagged signal events,
mistagged signal events, and background events. The parameters
$Y^{C}_{S}$ and $Y_{B}$ are the yields of correctly tagged signal
events and background events, respectively, and are free parameters in
the fit. The parameter $f^{M}$ is the fraction of signal events that
are mistagged and is determined from MC simulation. The signal mass
probability functions $S^{C}(m)$ and $S^{M}(m)$ are each the sum of
two Gaussian functions sharing the same mean, and describe the mass
distribution for correctly tagged and mistagged signal events,
respectively.

In the fit, the mean, the four Gaussian $\sigma$ parameters, and two
fractions relating the contribution of each Gaussian, are determined
from simulation, which has been found to accurately reproduce the
data. The function $S^a(\theta_K,\theta_\ell,\varphi)$ describes the
signal in the three-dimensional (3D) space of the angular variables
and corresponds to Eq.~(\ref{eq:PDF}).  The combination
$B^m(m) \, B^{\theta_K}(\theta_K) \, B^{\theta_\ell}(\theta_\ell) \,
B^{\varphi}(\varphi)$ is obtained from $\mathrm{B}^0$ sideband data
and describes the background in the space of
$(m,\theta_K,\theta_\ell,\varphi)$, where the mass distribution is an
exponential function and the angular distributions are polynomials
ranging from second to fourth degree, for both
$B^{\theta_K}(\theta_K)$ and $B^{\theta_\ell}(\theta_\ell)$, depending
on the $q^2$ bin, while the term $B^{\varphi}(\varphi)$ is of first
degree for all $q^2$ bins. The functions
$\epsilon^{C}(\theta_K,\theta_\ell,\varphi)$ and
$\epsilon^{M}(\theta_K,\theta_\ell,\varphi)$ are the efficiencies in
the 3D space of
$-1 \leq \cos\theta_K \leq 1, 0 \leq \cos\theta_\ell\leq 1$, and
$0\leq \varphi\leq \pi$ for correctly tagged and mistagged signal
events, respectively.

The fit is performed in two steps.  The initial fit uses the data from
the sidebands of the $\mathrm{B}^0$ mass to obtain the $B^m(m)$,
$B^{\theta_K}(\theta_K)$, $B^{\theta_\ell}(\theta_\ell)$, and
$B^{\varphi}(\varphi)$ distributions (the signal component is absent
from this fit). The sideband regions are
$3\sigma_{m} < |m-m_{\mathrm{B}^0}| < 5.5\sigma_{m}$, where $\sigma_m$
is the average mass resolution ($\approx$45MeV), obtained from fitting
the MC simulation signal to a sum of two Gaussians with a common
mean. The distributions obtained in this step are then fixed for the
second step, which is a fit to the data over the full mass range. The
free parameters in this fit are the angular parameters $P_1$, $P_5'$,
and $A^5_\mathrm{S}$, and the yields $Y^{C}_{S}$ and $Y_{B}$. In the
fits, the angular parameters $F_\mathrm{L}$, $F_\mathrm{S}$, and
$A_\mathrm{S}$ are fixed to previous CMS measurements performed on the
same data set with the same event selection
criteria~\cite{CMSKstarmumu2016}.

\begin{table}[!htb]
\centering
\caption{
Systematic uncertainties in $P_1$ and $P_5'$. For each source, the
range indicates the variation over the bins in $q^2$.}
\label{tab:systematics}
{
\begin{tabular}{lcccc}
Source & $P_1 (\times 10^{-3})$ & $P_5' (\times 10^{-3})$ \\
\hline \\[-2ex]
Simulation mismodeling       &   \x1--33   &  10--23  \\
Fit bias                     &   \x5--78   &  \x10--120 \\
Finite size of simulated samples  &  29--73   &  \x31--110 \\
Efficiency                   &  \x17--100  &   \x5--65  \\
$K\pi$ mistagging          &   \x\x8--110  &   \x6--66  \\
Background distribution      &  12--70   &  10--51  \\
Mass distribution            &      12   &      19  \\
Feed-through background      &   \x4--12   &   \x3--24  \\
$F_{\mathrm{L}}$, $F_{\mathrm{S}}$, $A_{\mathrm{S}}$ uncertainty propagation & \x\x0--210 & \x\x0--210 \\
Angular resolution           &   \x2--68   & 0.1--12\y  \\
\hline
Total                        & 100--230  &  \x70--250 \\
\end{tabular}
}
\end{table}

The fit formalism and results are validated through fits to
pseudo-experimental samples, MC simulation samples, and control
channels. To ensure correct coverage for the uncertainties of the
angular parameters, the Feldman-Cousins (FC)
method~\cite{Feldman_Cousins} is used with nuisance parameters. The
possible sources of systematic uncertainties investigated are
summarized in Table~\ref{tab:systematics}. For the $F_\mathrm{L}$,
$F_\mathrm{S}$, $A_\mathrm{S}$ uncertainty propagation, in the final
fit, we fix the parameters, $F_\mathrm{L}$, $F_\mathrm{S}$,
$A_\mathrm{S}$, at the previous CMS
measurements~\cite{CMSKstarmumu2016}. Their uncertainties are
propagated to the final results.

\begin{figure*}[!htb]
  \begin{center}
    \includegraphics[width=0.9\textwidth]{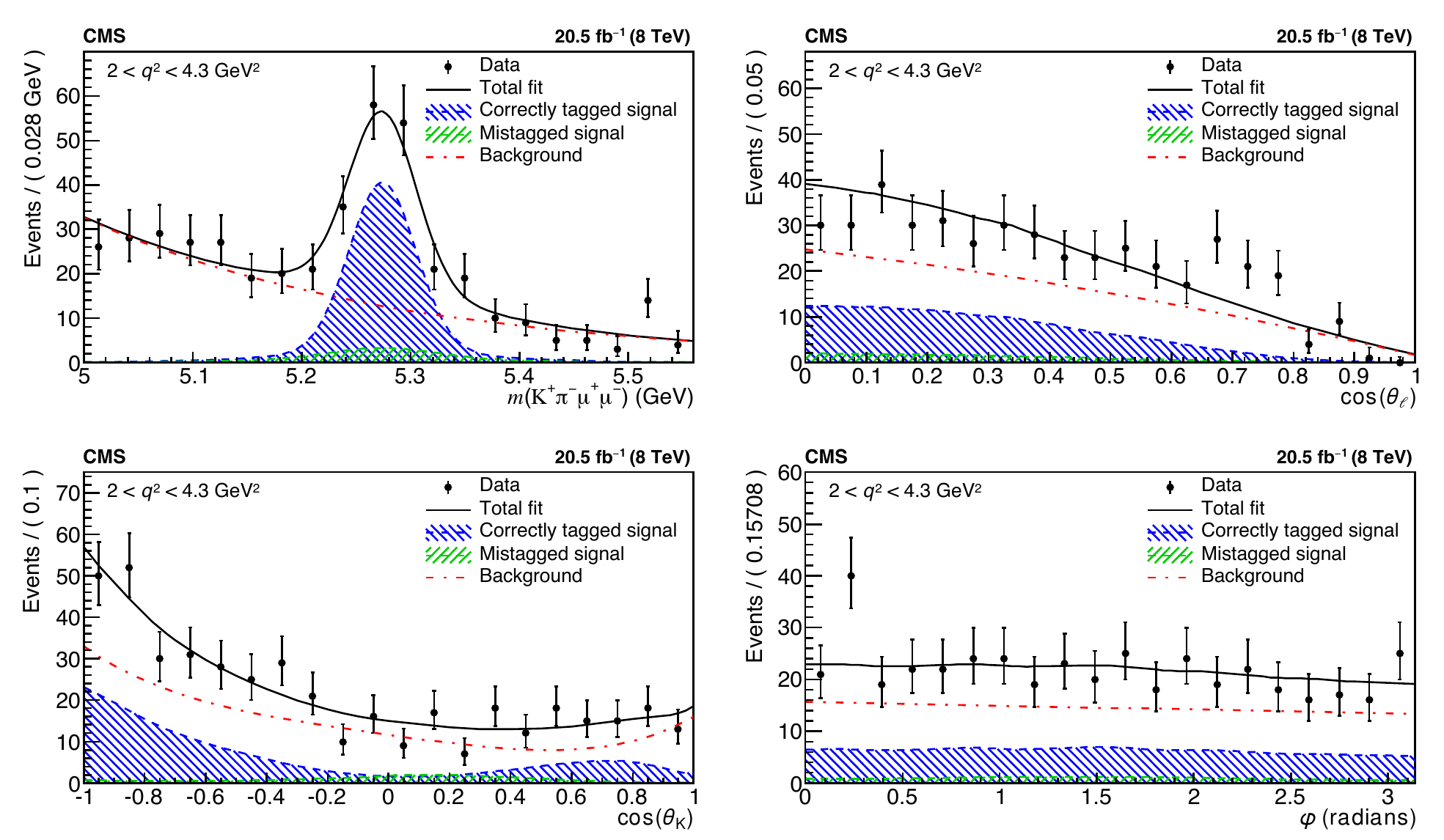}
    \caption{$\mathrm{K}^{+}\mathrm{\pi}^{-} \mu^ +\mu^-$ invariant
      mass and angular distributions for the second $q^2$ bin
      $2.00<q^2<4.30~\mathrm{GeV}^2$.  Overlaid on each plot is the
      projection of the results for the total fit, as well as for the
      three components: correctly tagged signal, mistagged signal, and
      background. The vertical bars indicate the statistical
      uncertainties~\cite{Sirunyan:2017dhj}.}
    \label{fig:fullplots}
  \end{center}
\end{figure*}

The signal data, corresponding to 1397 events, are fit in seven $q^2$
bins from 1 to $19~\mathrm{GeV}^2$. As an example, distributions for the
second $q^2$ bin, along with the fit projections, are shown in
Fig.~\ref{fig:fullplots}. 

\begin{figure}[!htb]
  \begin{center}
    \includegraphics[width=0.49\textwidth]{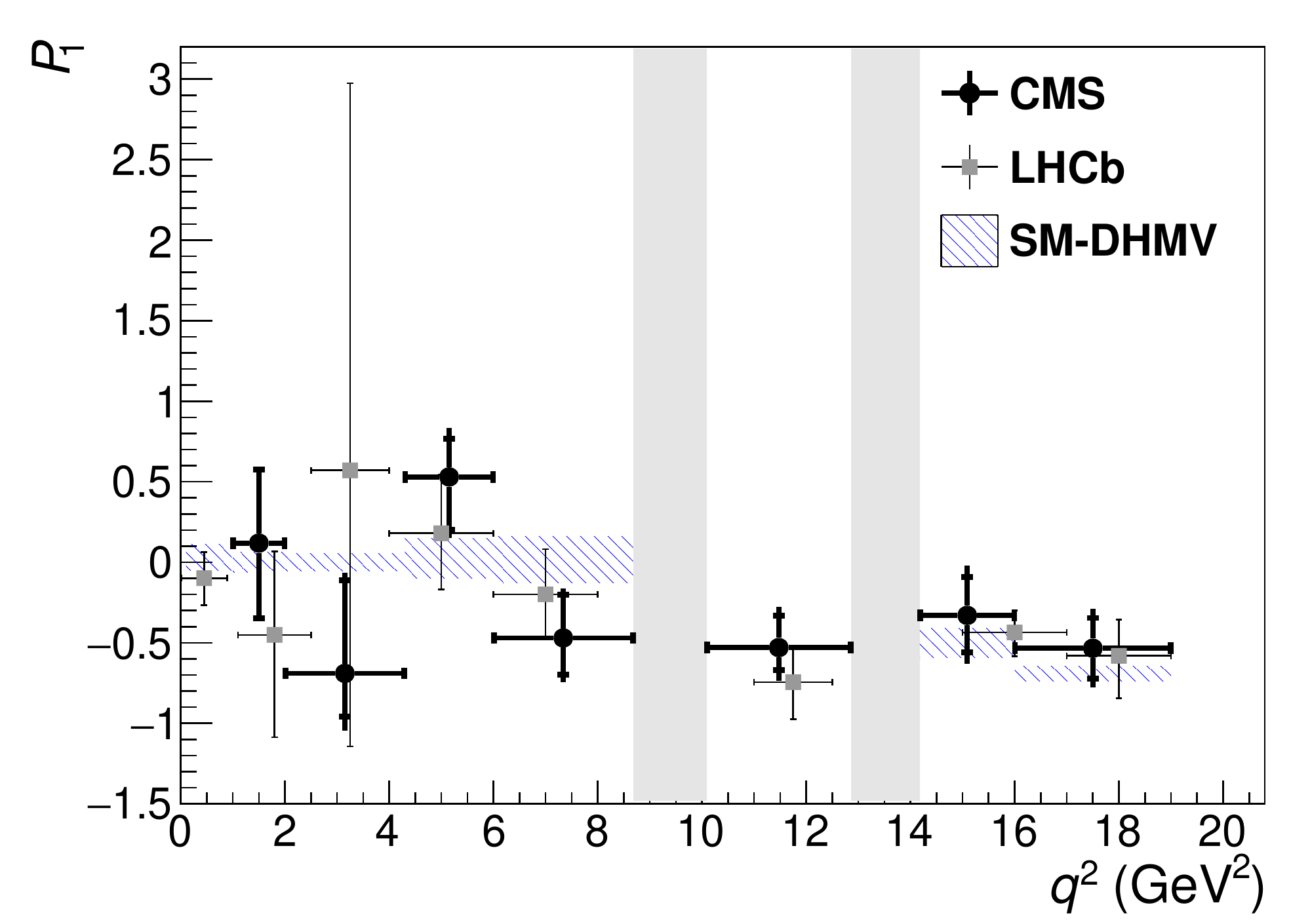}
    \includegraphics[width=0.49\textwidth]{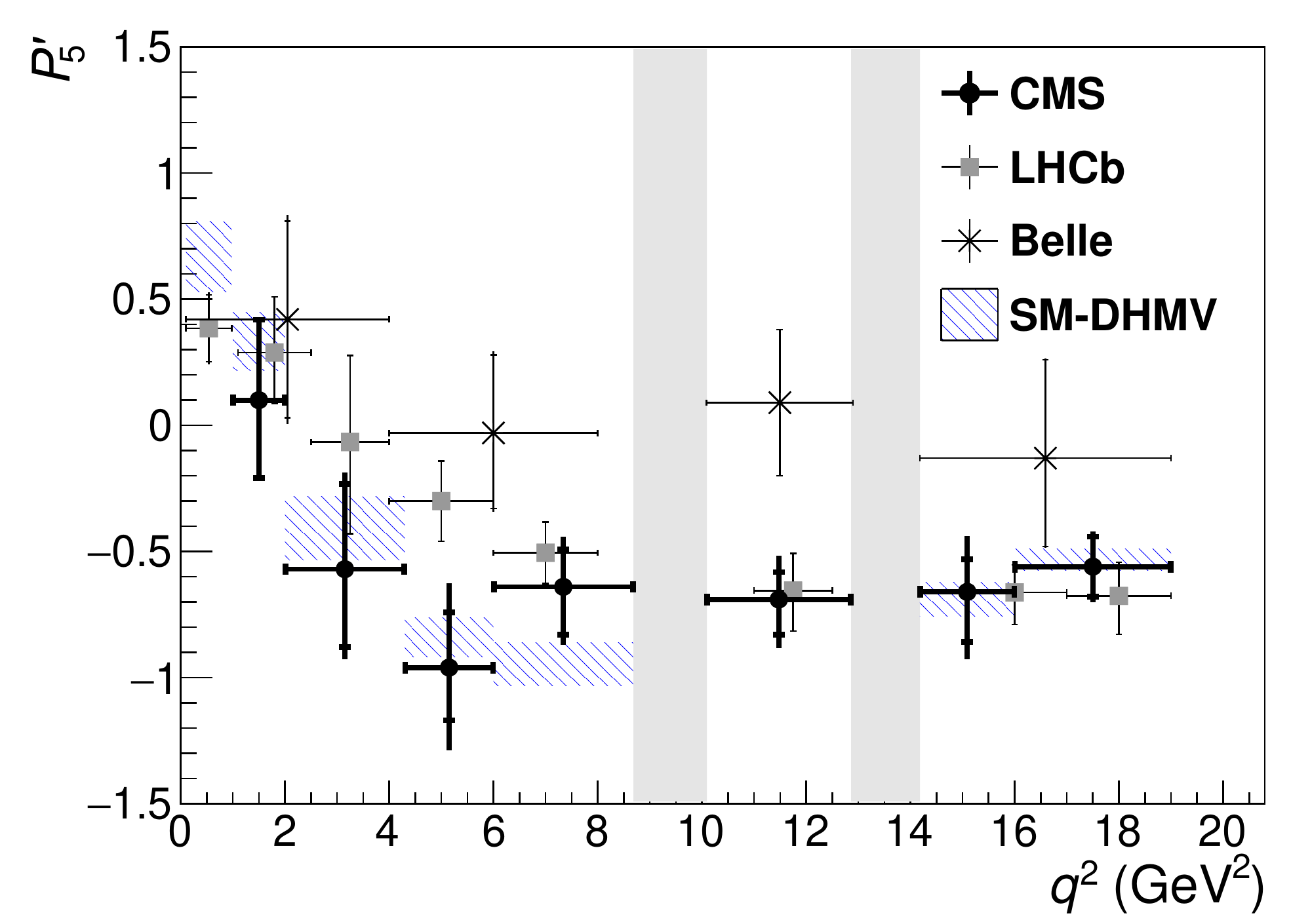}
    \caption{CMS measurements of the (left) $P_1$ and (right) $P_5'$
      angular parameters versus $q^2$ for \BtoKstmumu
      decays~\cite{Sirunyan:2017dhj}, in comparison to results from
      the LHCb~\cite{LHCbP5p2} and Belle~\cite{BelleP5p}
      Collaborations.  The statistical uncertainties are shown by the
      inner vertical bars, while the outer vertical bars give the
      total uncertainties.  The horizontal bars show the bin
      widths. The vertical shaded regions correspond to the $J/\psi$
      and $\psi'$ resonances.  The hatched region shows the prediction
      from SM calculations from
      Refs.~\cite{DescotesGenon:2012zf,Descotes-Genon:2013vna},
      averaged over each $q^2$ bin.}

    \label{fig:results}
  \end{center}
\end{figure}

The fitted values of $P_1$, and $P_5'$, along with their associated
uncertainties, for each of the $q^2$ regions are shown in
Fig.~\ref{fig:results}, along with the SM predictions. The results are
among the most precise to date for these parameters and are consistent
with predictions based on the standard model.

\section{Angular analysis of \Sig~\cite{Sirunyan:2018jll}}
\label{sec:kmumu}

The decay rate for the process \Sig depends on
$\cos\theta_{\ell}$, where $\theta_{\ell}$ is the angle between the
directions of the $\mu^-$ and ${\mathrm{K}}^+$ in the dilepton rest frame.
The $\cos\theta_{\ell}$ dependence of the decay width $\Gamma_{\ell}$ can be
parametrized~\cite{Theory:2007,ref2000,ref2001} in terms of the observables of
interest $A_{\mathrm{FB}}$ and $F_{\mathrm{H}}$ as:
\begin{equation} \label{eq-AngAFBFH}
  \frac{1}{\Gamma_{\ell}}\frac{\Gamma_{\ell}}{\cos\theta_\ell} = \frac{3}{4}(1-F_{\mathrm{H}})(1-\cos^{2}\theta_{\ell}) + \frac{1}{2}F_{\mathrm{H}} + A_{\mathrm{FB}}\cos\theta_{\ell}.
\end{equation}
The requirement for the decay rate to remain positive over all
possible lepton angles constrains the parameter space to the region
$0 \leq F_{\mathrm{H}} \leq 3$ and
$|A_{\mathrm{FB}}| \leq \text{min}(1, F_{\mathrm{H}}/2)$.

In the analysis, the selected events are reconstructed through the
decay into the fully charged final state of one charged hadron and a
pair of oppositely charged muons. Events from the control channels
\JPS and \PsP have the same final state as the signal process
${\rm B}{}^{+}$$\rightarrow$${\rm
  K{}^{+}}$$\mu^+\mu^-$.  Dimuon candidates are formed from two
oppositely charged muons matching the HLT criteria that triggered the
event readout. To discriminate signal events from background,
additional selection criteria on kinematic variables are used. These
selection criteria are determined through a maximization of the
expected signal significance using MC signal events and the surviving
data events in the final
$\mathrm{B}^+$ meson invariant mass fitting region,
5.1--5.6~$\mathrm{GeV}$. After applying the selection criteria, less
than 10\% of the selected events contain multiple
$\mathrm{B}^+$ candidates. In these events, only the candidate with
the highest $\mathrm{B}^+$ decay vertex fit probability is retained.

Events with a dimuon invariant mass ($q$) close to the $J/\psi$ or
$\psi'$ resonance region are rejected to remove this contamination
from the control channels. The $J/\psi$ and $\psi'$ resonance regions
are defined as
$m_{J/\psi}^{\mathrm{PDG}}-5\sigma_{q} < q < m_{J/\psi}+3\sigma_{q}$ and
$|q-m_{\psi'}^{\mathrm{PDG}}| < 5\sigma_{q}$, respectively, where
$\sigma_{q}$ is the calculated uncertainty in $q$, and the PDG
superscript indicates the world-average mass value~\cite{PDG} for each
particle. We further suppress such events by requiring,
$|(m-m_{\mathrm{B}^+}^{\mathrm{PDG}})-(q-m_{J/\psi}^{\mathrm{PDG}})|
> 0.13~\mathrm{GeV}$ and
$|(m-m_{\mathrm{B}^+}^{\mathrm{PDG}})-(q-m_{\psi'}^{\mathrm{PDG}})|
> 0.06~\mathrm{GeV}$ in the $\mathrm{B}^+$ meson invariant mass region of
5.1--5.6~$\mathrm{GeV}$.

MC simulated event samples are widely used in
the analysis. The number of simulated events for the signal sample
\Sig corresponds to more than 160 times that of the data. The control
channels for this analysis are \JPS, \PsP, where the muon pairs come
from $J/\psi$ or $\psi'$ decays.

The angular
observables $A_{\mathrm{FB}}$ and $F_{\mathrm{H}}$ are extracted from
a two-dimensional extended unbinned maximum-likelihood fit to the
angular distribution of the selected $\mathrm{B}^+$ meson candidates in
each $q^{2}$ range. The unnormalized probability density function
(pdf) used in the two-dimensional fit is:
\begin{equation} \label{eq:PDF-k}
      \text{pdf}(m, \cos\theta_{\ell}) = Y_{\mathrm{S}} \times S_{\mathrm{m}}(m) \times S_{\mathrm{a}}(\cos\theta_{\ell}) \times \epsilon(\cos\theta_{\ell}) + Y_{\mathrm{B}} \times B_{\mathrm{m}}(m) \times B_{\mathrm{a}}(\cos\theta_{\ell}),
\end{equation}
where the two contributions on the righthand side correspond to the
parametrization of the signal and background. The parameters $Y_{\mathrm{S}}$
and $Y_{\mathrm{B}}$ are the yields of signal and background events,
respectively. The functions $S_{\mathrm{m}}(m)$ and
$S_{\mathrm{a}}(\cos\theta_{\ell})$ describe the signal invariant mass
and angular distributions, while $B_{\mathrm{m}}(m)$ and
$B_{\mathrm{a}}(\cos\theta_{\ell})$ are similar functions describing
the background. The function $\epsilon(\cos\theta_{\ell})$ is the
signal efficiency as a function of $\cos\theta_{\ell}$.

The signal efficiency $\epsilon(\cos\theta_{\ell})$ is factorized into
an acceptance $\epsilon_{\text{acc}}$ times a reconstruction
efficiency $\epsilon_{\text{reco}}$, which are both functions of
$\cos\theta_{\ell}$.  The signal efficiency
$\epsilon(\cos\theta_{\ell})$ is parametrized and fit with a
sixth-order polynomial for the nine different signal $q^2$ ranges used
in this analysis. The signal distribution $S_{\mathrm{m}}(m)$ is
modeled as the sum of two Gaussian functions with a common mean, and
$S_{\mathrm{a}}(\cos\theta_{\ell})$ is given in
Eq.~(\ref{eq-AngAFBFH}). The background distribution
$B_{\mathrm{m}}(m)$ is modeled as a single exponential function, while
$B_{\mathrm{a}}(\cos\theta_{\ell})$ is parametrized as the sum of a
Gaussian function and a third- or fourth-degree polynomial, depending
on the particular $q^2$ range. Many of the parameters in the final fit
are set to a given value with a Gaussian constraint that reflects the
input uncertainty of the value.  The free parameters of the fit are
$Y_{\mathrm{S}}$, $Y_{\mathrm{B}}$, $A_{\mathrm{FB}}$, and
$F_{\mathrm{H}}$, as well as the exponential decay parameter of
$B_{\mathrm{m}}(m)$.

To validate the efficiency description derived from simulation, we
check that the ratio of the branching fractions of the two control
channels is consistent with the world-average value~\cite{PDG} within
their uncertainties.  The MC simulation samples are used to validate
the fitting procedure in each $q^2$ range.

Several sources of systematic uncertainties are considered in this analysis.
All systematic uncertainties in the measured values of $A_{\mathrm{FB}}$ and
$F_{\mathrm{H}}$, and the total systematic uncertainties are summarized in Table~\ref{tab:sysAFBFH}.

\begin{table}[!htb]
    \centering
	\caption{Absolute values of the uncertainty contributions in the measurement
      of $A_{\mathrm{FB}}$ and $F_{\mathrm{H}}$. For
      each item, the range indicates the variation of the uncertainty
      in the signal $q^2$ bins.}
	 \begin{tabular}{lcc}
           Systematic uncertainty & $A_{\mathrm{FB}}~(\times10^{-2})$ & $F_{\mathrm{H}}~(\times10^{-2})$ \\
           \hline
           Finite size of MC samples         & 0.4--1.8 & 0.9--5.0 \\
           Efficiency description            & 0.1--1.5 & 0.1--7.8 \\
           Kinematic mismodeling             & 0.1--2.8 & 0.1--1.4 \\
           Background parametrization model  & 0.1--1.0 & 0.1--5.1 \\
           Angular resolution                & 0.1--1.7 & 0.1--3.3 \\
           Dimuon mass resolution            & 0.1--1.0 & 0.1--1.5 \\
           Fitting biases                    & 0.1--3.2 & 0.4--25 \\
           Background distribution           & 0.1--7.2 & 0.1--29 \\
           \hline
           Total systematic uncertainty      & 1.6--7.5 & 4.4--39 \\
    \end{tabular}
	\label{tab:sysAFBFH}
\end{table}

\begin{figure}[!hbt]
  \begin{center}
    \includegraphics[width=0.49\textwidth]{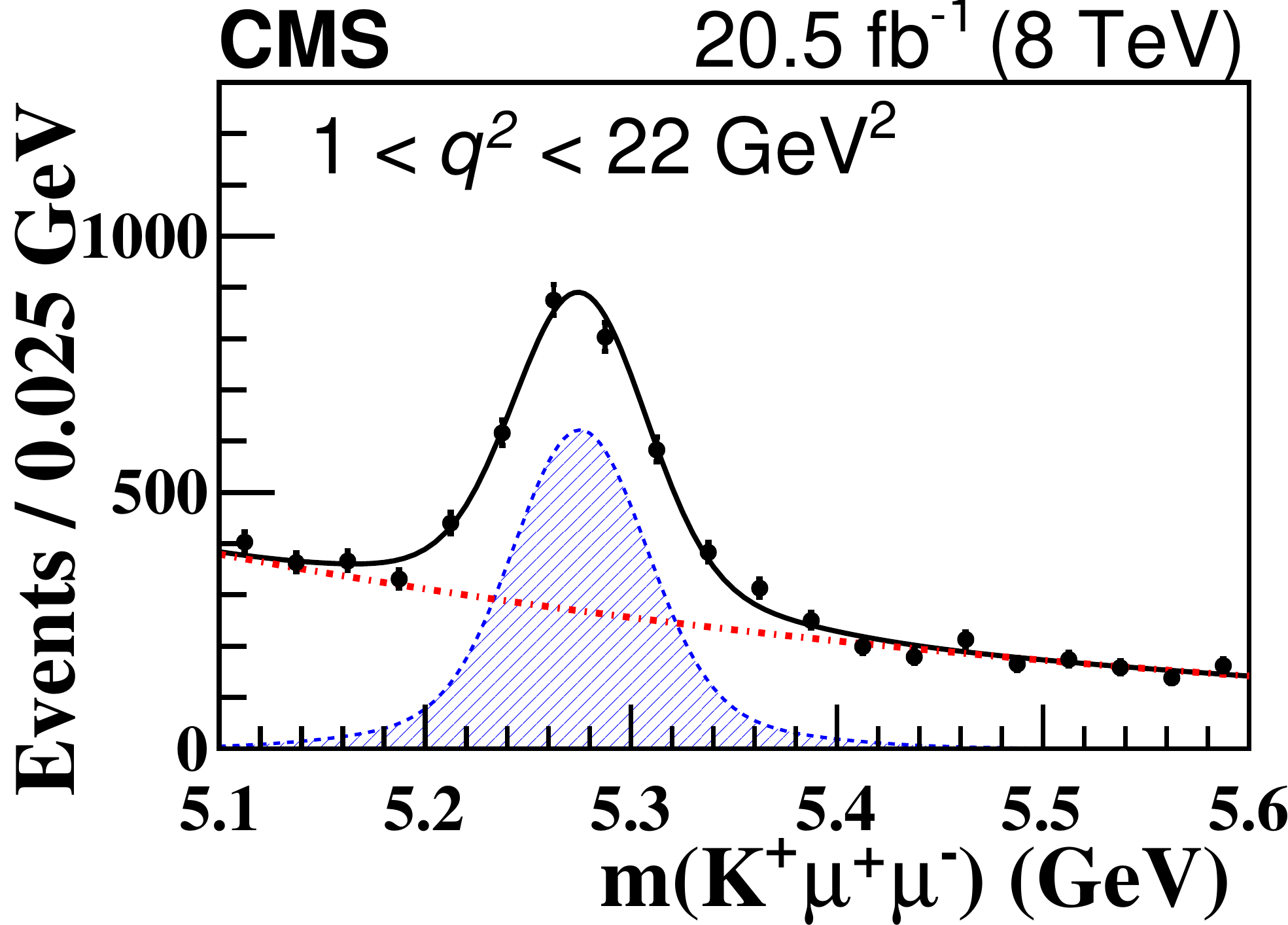}
    \includegraphics[width=0.49\textwidth]{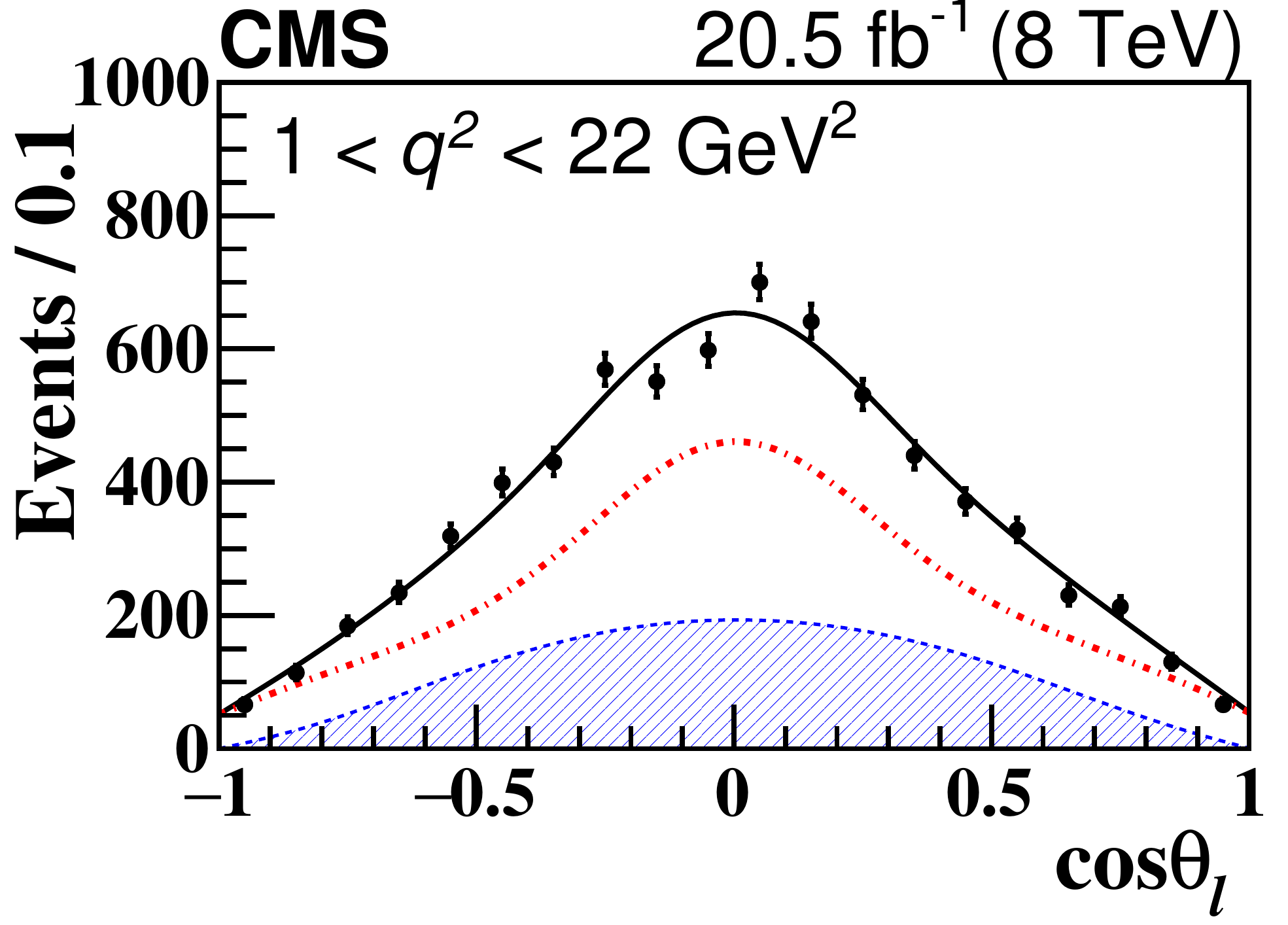}
    \caption{Projections of the fit results from data for the
      ${\rm K}^+\mu^+\mu^-$ invariant mass and $\cos(\theta_{\ell})$
      distributions for the inclusive $q^2$ bin of 1--22$~\mathrm{GeV}^2$
      (excluding the resonance regions). The vertical bars represent
      the statistical uncertainties in the data~\cite{Sirunyan:2018jll}.}
    \label{fig:kpproj}
    \end{center}
\end{figure}

The final fit is performed over the full $\mathrm{B}^+$ meson
invariant mass range and results in $2286\pm73$ signal events with
$q^2$ from 1 to 22$~\mathrm{GeV}^2$. Projections of the fit results from data
for the ${\rm K}^+\mu^+\mu^-$ invariant mass and $\cos(\theta_{\ell})$
distributions for the inclusive $q^2$ bin of 1--22$~\mathrm{GeV}^2$ (excluding
the resonance regions) are shown in Fig.~\ref{fig:kpproj}. To evaluate
the statistical uncertainties, the 68.3\% confidence level intervals
on $A_{\mathrm{FB}}$ and $F_{\mathrm{H}}$ are estimated using the
profiled Feldman--Cousins technique~\cite{Feldman_Cousins}. The
systematic and statistical uncertainties are added in quadrature to
obtain the total uncertainty.

\begin{figure}[!htb]
  \begin{center}
    \includegraphics[width=0.49\textwidth]{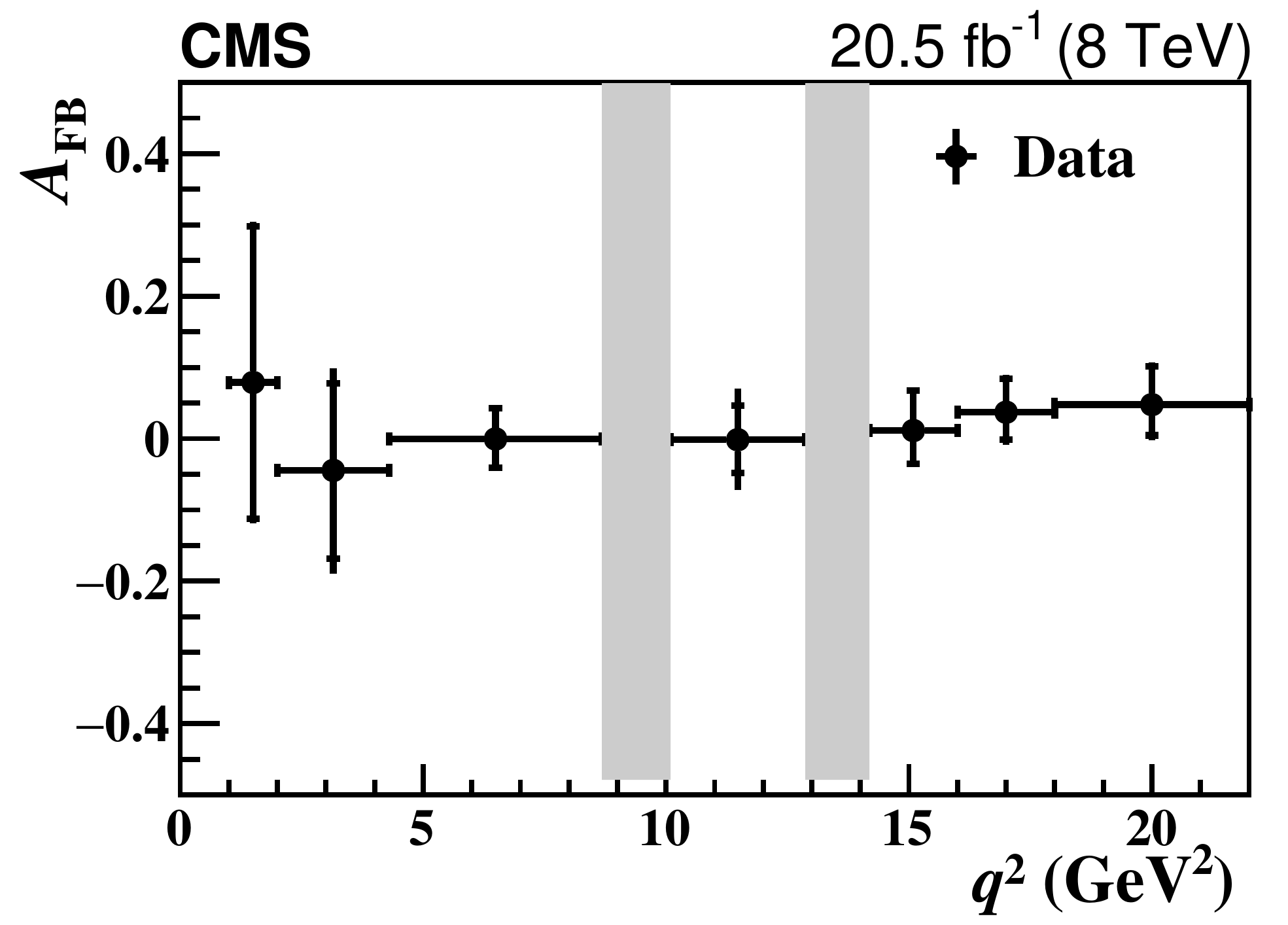}
    \includegraphics[width=0.49\textwidth]{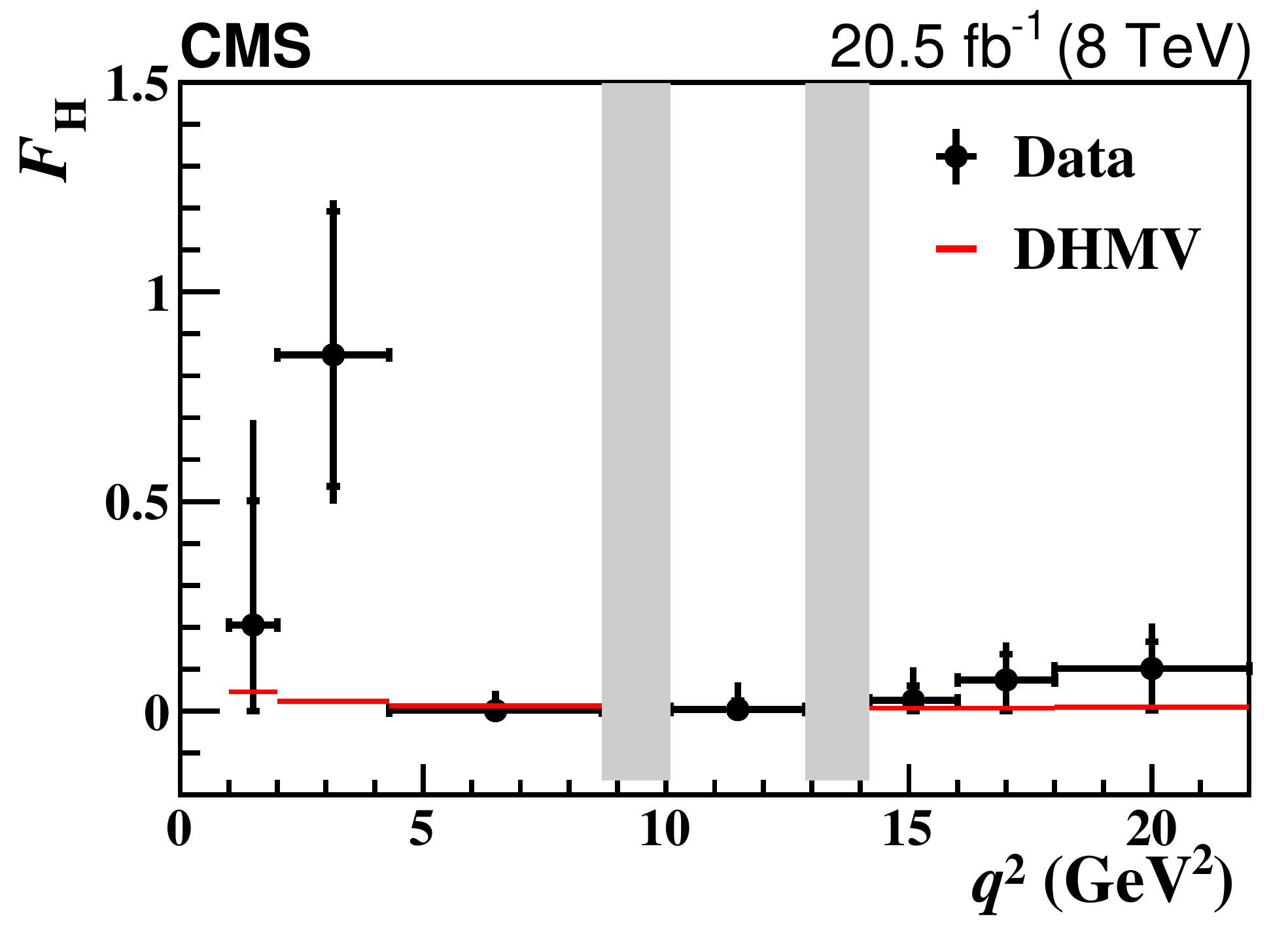}
    \caption{Results of the measurement of $A_{\text{FB}}$ (left) and
      $F_{\text{H}}$ (right) in bins of $q^2$~\cite{Sirunyan:2018jll}. The statistical
      uncertainties are shown by the inner vertical bars, while the
      outer vertical bars give the total uncertainties.  The
      horizontal bars show the bin widths.  The vertical shaded
      regions are 8.68--10.09 and 12.86--14.18$~\mathrm{GeV}^2$, and
      correspond to the $\mathrm{J}/\psi$ and $\psi'$-dominated
      control regions, respectively.  The red line in the right plot
      shows the DHMV SM theoretical
      prediction~\cite{Descotes-Genon:2014uoa,Descotes-Genon:2015uva}.
	  %%The predictions from two other calculations are given in Table~\ref{tab:resFHAFB}.
    }
    \label{fig:resultFHAFB}
  \end{center}
\end{figure}

The measured values of $A_{\mathrm{FB}}$ and $F_{\mathrm{H}}$ for each
$q^2$ range are shown in Fig.~\ref{fig:resultFHAFB}. The measured
values of $A_{\mathrm{FB}}$ are consistent with the SM expectation of
no asymmetry. We also compare the measured results with three SM
predictions for $F_{\mathrm{H}}$ with different input parameters and
different handling of higher-order corrections, one of which is also
shown in Fig.~\ref{fig:resultFHAFB}. There is generally good agreement
between the predictions and our results, as well as between our
results and previous
measurements~\cite{Babar2006,Belle2009,CDF2012,LHCb2012An,LHCb2014An}.

\section{Summary and outlook}

Using pp collision data recorded at $\sqrt{s}=8~\mathrm{TeV}$ with the
CMS detector at the LHC, corresponding to an integrated luminosity of
$20.5~\mathrm{fb}^{-1} $, angular analyses have been performed for the
decays of $\mathrm{B}^0 \to \mathrm{K}^{*0} \mu^ +\mu^-$ and
$\mathrm{B}^{+}$$\to$$\mathrm{K}^{+}$$\mu^+\mu^-$. For each bin of the
dimuon invariant mass squared $(q^2)$, unbinned maximum-likelihood
fits were performed to the distributions of the B meson invariant mass
and the three decay angles, to obtain values of angular
parameters. The results are among the most precise to date and are
consistent with previous measurements and with standard model
predictions.  CMS efforts with more channels and more coming data will
be continued to further test the standard model with higher precision
in future.

% \bibliographystyle{h-physrev}
% \bibliography{wangdy}

\end{document}